\documentclass[english,aps,prd,superscriptaddress,preprint]{revtex4}
\usepackage[latin9]{inputenc}
\usepackage{graphicx}

\providecommand{\tabularnewline}{\\}

\makeatother

\usepackage{babel}

\begin{document}
\draft
\title{Single Production of Fourth Family $t'$ Quarks at LHeC}

\author{O. Cakir}
\email[ ocakir@science.ankara.edu.tr]{}
\affiliation{Ankara
University, Faculty of Sciences, Department of Physics, 06100,
Tandogan, Ankara, Turkey}

\author{A. Senol}
\email[asenol@kastamonu.edu.tr]{}
\affiliation{Kastamonu University,
Faculty of Arts and Sciences, Department of Physics, 37100,
Kuzeykent, Kastamonu, Turkey}

\author{A. T. Tasci}
\email[atasci@kastamonu.edu.tr]{}
\affiliation{Kastamonu University,
Faculty of Arts and Sciences, Department of Physics, 37100,
Kuzeykent, Kastamonu, Turkey}

\begin{abstract}
We study the single production of fourth-family $t'$ quarks via the
process $ep\to t'\nu$ at Large Hadron electron Collider (LHeC). We
calculate the background and signal cross sections for the mass range
300-800 GeV. It is shown that the LHeC can discover single $t'$ quark
up to the mass of $800$ GeV for the optimized mixing parameters.
\end{abstract}
\maketitle

\section{Introduction}

The start-up of the Large Hadron Collider (LHC) will scan a new
range of energy and mass to solve the well known unanswered
questions in particle physics; especially concerned, two of them are
flavor problem and electroweak symmetry breaking. The unification of
these two problems is the major motivation to consider fourth
family. In addition to the LHC, $ep$ facility will also provide
complementary information for new physics beyond the Standard Model
(SM). The Large Hadron electron Collider (LHeC)
\cite{Dainton:2006wd} will consist of a new linear accelerator or a
storage ring to collide electrons with energy of 70/140 GeV with the
existing 7 TeV proton beam from the LHC. This will result in deep
inelastic scattering interactions with a center of mass energy of
1.4/1.9 TeV, and with a luminosity of up to $10^{33}/10^{32}$
cm$^{-2}$s$^{-1}$, both significantly greater than the only previous
electron-proton collider at HERA. The LHeC will also be expected to
have sensitivity to new physics and new states of matter. It would
give a possibility for polarized $ep$ scattering to investigate the
couplings of new fermions to the Standard Model (SM) fermions.

It is well known that the SM does not predict the number of fermion
families. The fermion family replicants remain a mystery of the
flavor problem. The discovery of a sequential fourth family may play
an important role in understanding the flavor structure of the SM.
Determination of the number of fermion families will be an important
goal of the upcoming experiments at the LHC
\cite{Holdom:2006mr,Holdom:2007nw,Holdom:2007ap,Kribs:2007nz,Ozcan:2008zz,Cakir:2008su},
and further at the ILC and CLIC \cite{Ciftci05}. Meanwhile, the
production of fourth family quarks via flavor changing neutral
current interactions are investigated at future colliders
\cite{Arik:2002sg,Arik:2003vn,Alan:2003za,Arhrib:2006pm,Herrera:2008yf}.

The experiments at Tevatron have already constrained the masses of
fourth family quarks. The collider detector at Fermilab (CDF) has
searched strong pair production of $t'$ with its associated anti-quark,
each decaying to a $W$ boson and a jet, with 2.8 $fb^{-1}$ of data
Run II of Tevatron setting a lower bound on mass of $t'$ quark, $m_{t'}>311$
GeV at $95\%$ CL \cite{Lister:2008is}. There seems to be some parameter
space (mass vs. mixing angle) of the fourth family quarks which could
be explored at future searches \cite{Holdom:2006mr,Holdom:2007nw,Holdom:2007ap,Herrera:2008yf,Cakir:2008su,Kribs:2007nz}.

In this study, we investigate the discovery potential of the LHeC
for the single production of sequential fourth family $t'$ quarks
via the process $e^{+}p\to
t'\bar{\nu}_{e}(e^{-}p\to\bar{t'}\nu_{e})$ with the 70 GeV electron
and 7 TeV proton beam energies. We have calculated the cross
sections of signals and corresponding backgrounds. The decay widths
and branching ratios of $t'$ quark are calculated in the mass range
300-800 GeV. All the calculations have been performed with CompHEP
\cite{Pukhov:1999gg} by including the new interaction vertices.

\section{Single Production and decay of $t'$ quark}

The extension of the SM is simply to add a sequential fourth family
fermions where left-handed components transform as a doublet of
$SU(2)_{L}$ and right-handed components as singlets. The interaction
of the fourth family $t'$ quark with the quarks $q_{i}$ via the SM
gauge bosons ($\gamma,g,Z^{0},W^{\pm}$) is given by

\begin{eqnarray}
L & = & -g_{e}Q_{t}\overline{t}'\gamma^{\mu}t'A_{\mu}\nonumber \\
 &  & -g_{s}\overline{t}'T^{a}\gamma^{\mu}t'G_{\mu}^{a}\nonumber \\
 &  & -\frac{g}{2\cos\theta_{W}}\overline{t}'\gamma^{\mu}(g_{V}-g_{A}\gamma^{5})t'Z_{\mu}^{0}\nonumber \\
 &  & -\frac{g}{2\sqrt{2}}V_{t'q_{i}}\overline{t}'\gamma^{\mu}(1-\gamma^{5})q_{i}W_{\mu}^{\pm}+h.c.\label{eq:1}
 \end{eqnarray}
where $g_{e}$, $g$ are the electro-weak coupling constants, and
$g_{s}$ is the strong coupling constant. $A_{\mu}$, $G_{\mu}$,
$Z_{\mu}^{0}$ and $W_{\mu}^{\pm}$ are the vector fields for photon,
gluon, $Z^{0}$-boson and $W^{\pm}$-boson, respectively. $Q_{t'}$ is
the electric charge of fourth family quark $t'$; $T^{a}$ are the
Gell-Mann matrices. $g_{V}$ and $g_{A}$ are the vector and
axial-vector type couplings of the neutral weak current with $t'$
quark. Finally, the $V_{t'q}$ denotes the elements of extended
4$\times$4 CKM mixing matrix which are constrained by flavor
physics. In this study, we use the parametrization
\cite{Ozcan:2008zz}, $V_{t'd}$=0.063,$V_{t's}$=0.46,$V_{t'b}$=0.47,
which were optimized for a $1\sigma$ deviation over the average
values of the CKM matrix elements.

\begin{figure}[htbp!]
\centering\includegraphics[scale=0.6]{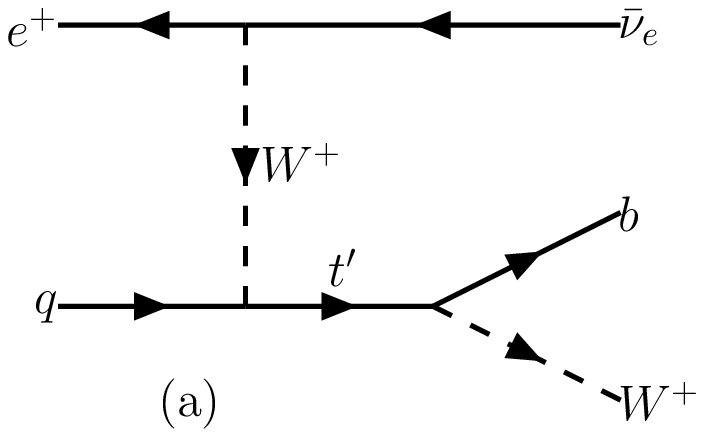} \includegraphics[scale=0.6]{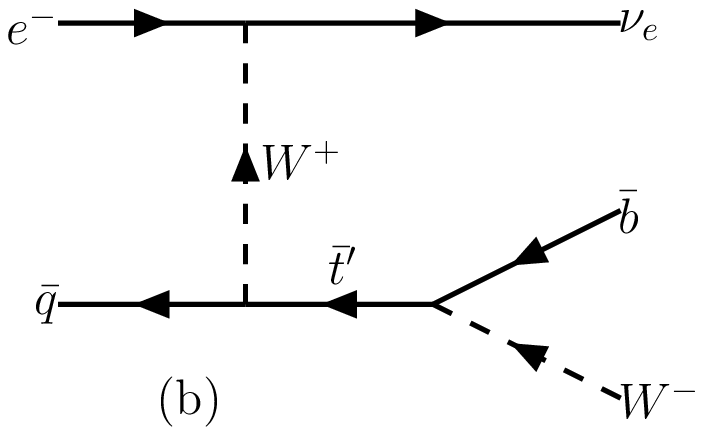}

\caption{The diagrams relevant for single production of fourth family $t'(\bar{t'})$
quark at LHeC\label{fig1}}

\end{figure}

\begin{figure}[htbp!]
\includegraphics[scale=0.6]{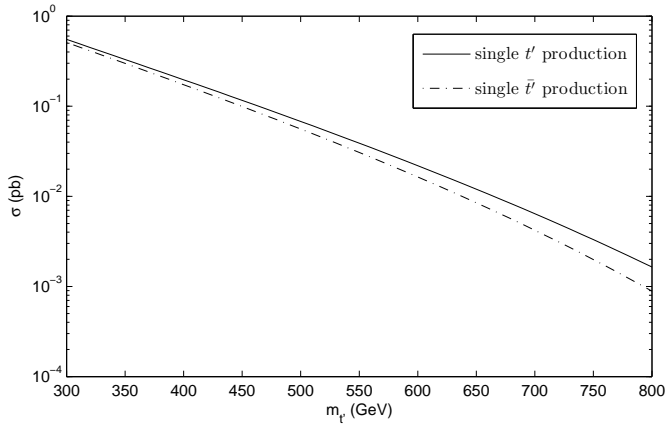}

\caption{The total cross sections at the LHeC for the processes $e^{+}p\to t'\bar{\nu}_{e}$
(solid line) and $e^{-}p\to\bar{t'}\nu_{e}$(dot-dashed line) with
$\sqrt{s}=1.4$ TeV.\label{fig2}}

\end{figure}

The tree level Feynman diagrams for the single production of
$t'$($\bar{t'}$) quark and their subsequent decays are given in Fig.
\ref{fig1}. The total decay widths of $t'$ quark within the SM
framework are presented in Table \ref{tab1} assuming the mass of
$t'$ quark between 300 and 800 GeV. For this parametrization $t'$
branchings remain unchanged $51\%(W^{+}b)$, $48\%(W^{+}s)$,
$0.9\%(W^{+}d)$ in the considered mass range. In Fig. \ref{fig2}, we
display the single production tree level cross-sections of the
fourth generation $t'$ (solid line) and $\bar{t'}$ (dot-dashed line)
quarks depending on their masses at the LHeC with $\sqrt{s}=1.4$
TeV. The cross sections of $t'$ and $\bar{t'}$ quarks does not
change significantly as seen in Fig. \ref{fig2}. Therefore, these
two processes will be considered in our following analysis. We use
the CTEQ6M \cite{Pumplin:2002vw} parton distribution function in our
numerical calculations.

\begin{table}[htbp!]
\caption{The total decay widths of $t'$ quark depending on its mass values.\label{tab1} }

\centering

\begin{tabular}{lll}
 &  & \tabularnewline
\hline
Mass (GeV)  &  & $\Gamma$(GeV) \tabularnewline
\hline
300  &  & 3.84\tabularnewline
400  &  & 9.19\tabularnewline
500  &  & 18.00\tabularnewline
600  &  & 31.14 \tabularnewline
700  &  & 49.48 \tabularnewline
800  &  & 73.87 \tabularnewline
\hline
\end{tabular}
\end{table}

\begin{figure}[htbp!]
\includegraphics[scale=0.6]{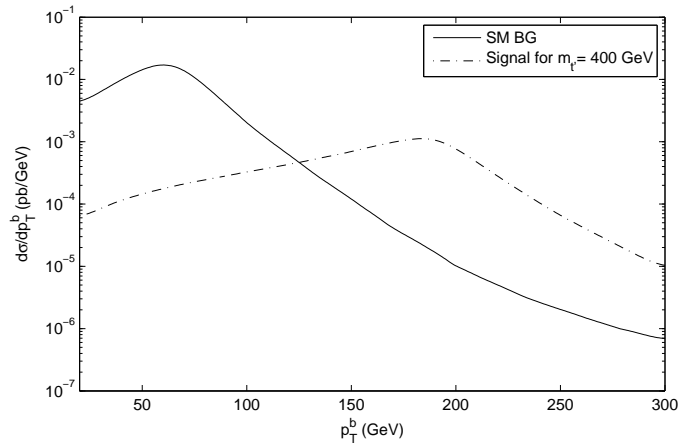}

\caption{The differential cross section depending on the transverse momentum
of the final state $b$ quark for the subprocess $e^{+}p\to W^{+}b\bar{\nu_{e}}$.
The solid line and dashed line correspond to the SM background and
the signal for $m_{t'}$=400 GeV, respectively.\label{fig3}}

\end{figure}

\begin{figure}[htbp!]
\includegraphics[scale=0.6]{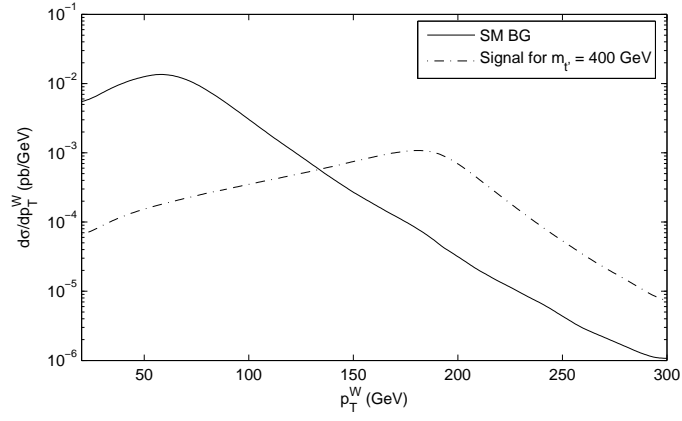}

\caption{The differential cross section depending on the transverse momentum
of the final state $W^{+}$-boson for the subprocess $e^{+}p\to W^{+}b\bar{\nu_{e}}$
at the LHeC. The solid line and dashed line correspond to the SM background
and signal for $m_{t'}$=400 GeV, respectively.\label{fig4}}

\end{figure}

\begin{figure}[htbp!]
\includegraphics[scale=0.6]{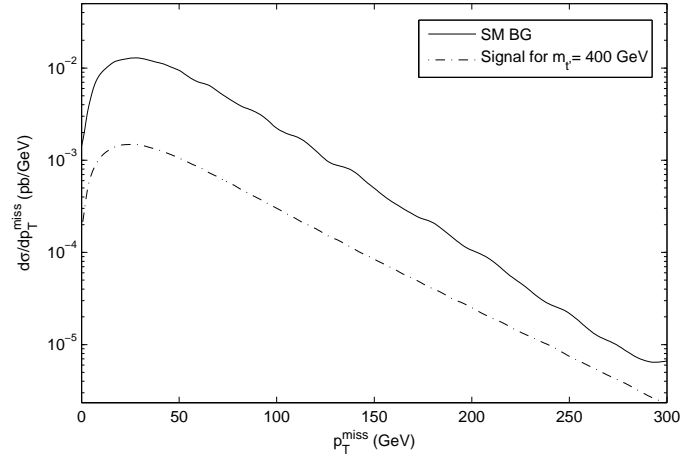}

\caption{The differential cross section depending on the transverse momentum
of the final state neutrino $\nu$ for the subprocess $ep\to Wb\nu$.
The solid line and dashed line correspond to the SM background and
signal for $m_{t'}$=400 GeV of this subprocess, respectively.\label{fig5}}

\end{figure}

The transverse momentum distributions of the final state $b$-quark
for the signal and background are shown in Fig. \ref{fig3}. The
distribution of the final state b-quark from $t'$ decays is analyzed
since b-quarks contribute substantially to the production mechanism.
For a single $t'$ quark production with a mass of 400 GeV, the
$p_{T}$ distribution shows a peak around 200 GeV, comparing this
distribution with that of the corresponding background we could
apply a $p_{T}$ cut to reduce the background. For the final state
$W^{+}b\nu_{e}$ we also plot the $p_{T}$ distributions of
$W^{+}$-boson and missing $p_{T}$ for neutrino as shown in Figs.
\ref{fig4} and \ref{fig5}. As can be seen from Fig. \ref{fig5}, a
cut on the missing transverse momentum $p_{T}^{miss}>50$ GeV is
required for the analysis.

\begin{figure}[htbp!]
\includegraphics[scale=0.6]{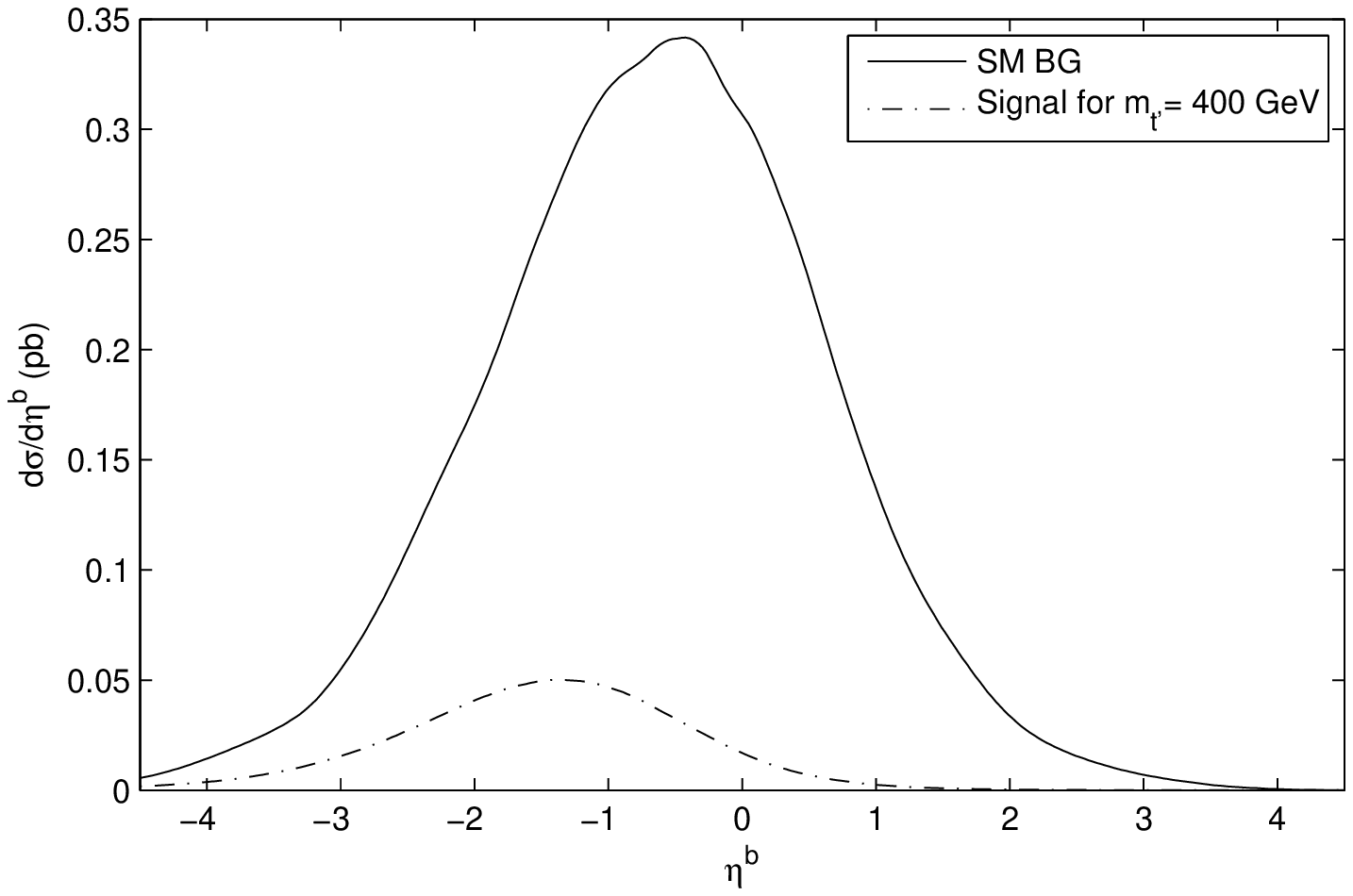}

\caption{The rapidity distribution of final state b quark of the subprocess
$e^{+}p\to W^{+}b\bar{\nu}_{e}$ for the SM background (solid line)
and signal with $m_{t'}$=400 GeV (dot-dashed line). \label{fig6}}

\end{figure}

\begin{figure}[htbp!]
\includegraphics[scale=0.6]{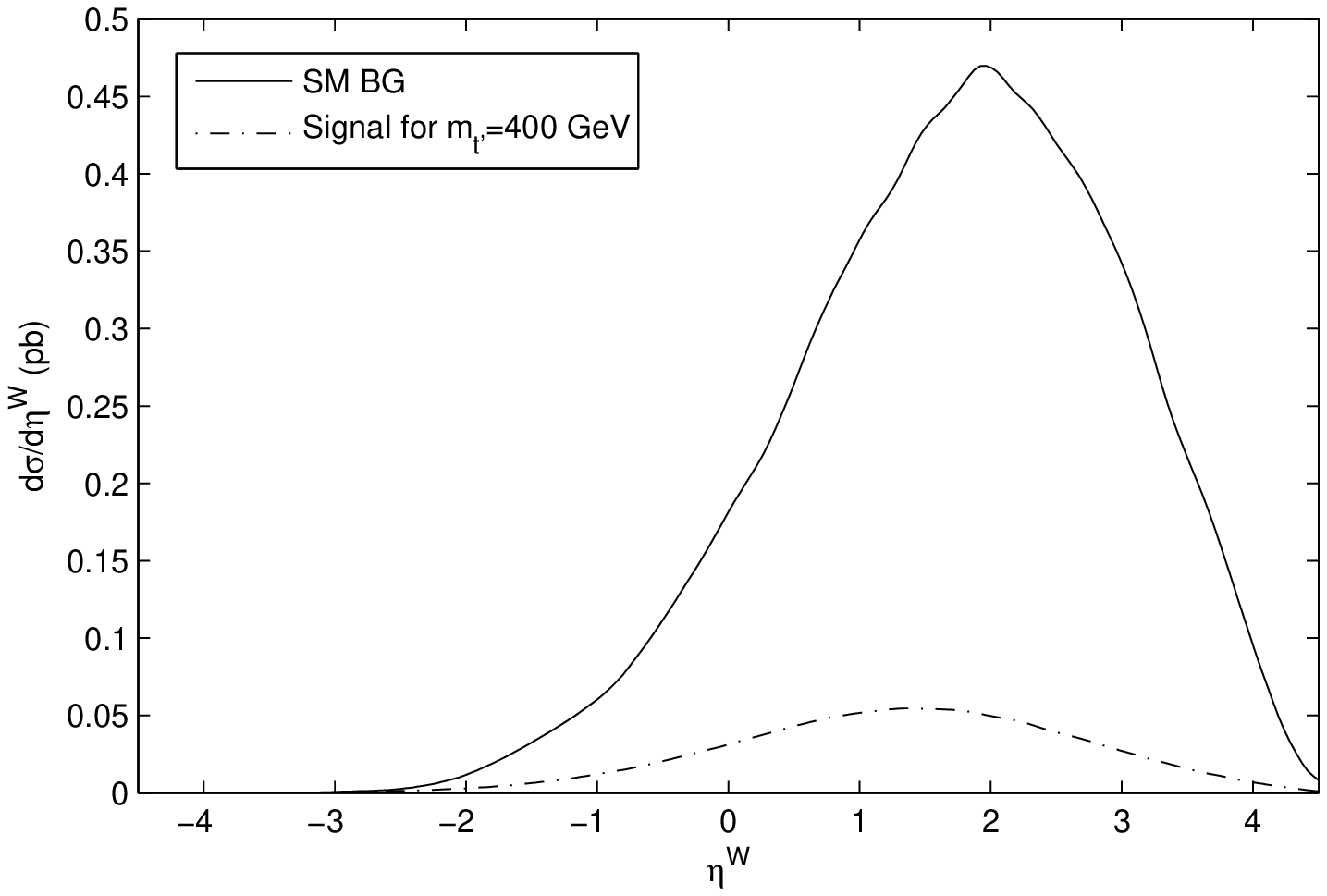}

\caption{The rapidity distribution of final state $W^{+}$-boson of the subprocess
$e^{+}p\to W^{+}b\bar{\nu}_{e}$ for the SM background (solid line)
and signal with $m_{t'}$=400 GeV (dot-dashed line).\label{fig7}}

\end{figure}

\begin{figure}
\includegraphics[scale=0.6]{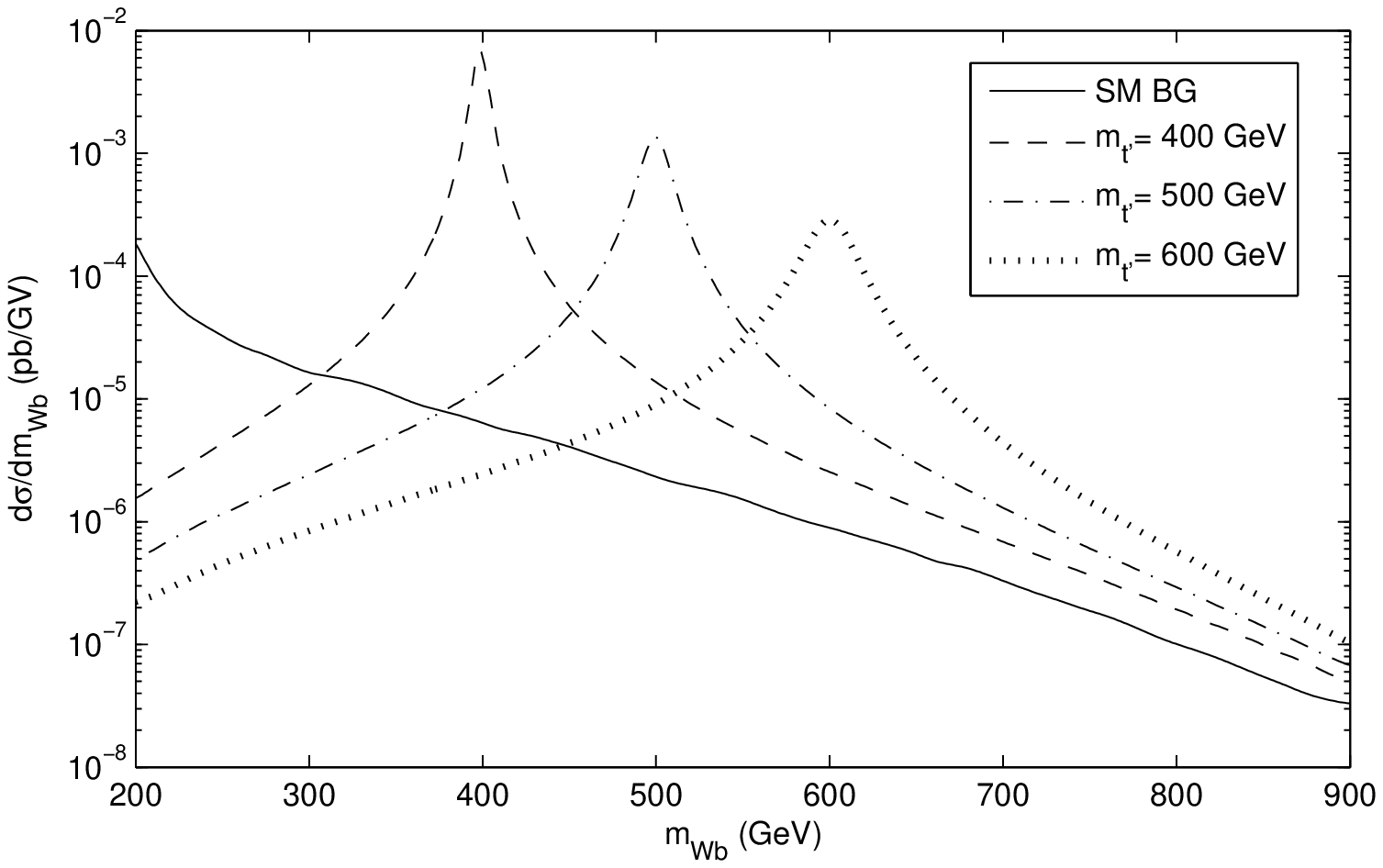}

\caption{The invariant mass distribution for the SM background (solid line)
and the $Wb$ signal from $t'$ decay for $m_{t'}$=400 GeV (dashed
line), 500 GeV(dot-dashed line) and 600 GeV (dotted line). \label{fig8}}

\end{figure}

The signal and background events will show different rapidity distributions
for the $b$-jet and $W^{\pm}$ boson in the final state as seen in
Figs. \ref{fig6} and \ref{fig7}. The invariant mass of $W^{+}b$
system is shown in Fig. \ref{fig8}. The peaks show the $t'$ signal
with the mass of 400, 500 and 600 GeV.

\section{Analysis}

In order to obtain the signal visible over the background we also
apply an invariant mass cut $|m_{t'}-m_{W^{+}b}|<10-20$ GeV
according to the mass and the decay width of $t'$ quark. Hence, we
obtain a significant reduction on the cross section of the
background. The background comes from SM events which yield exactly
the same final state particles as the signal process. Here we
consider the signal as a $b$-jet, the missing transverse energy
(MET), and a $W^{+}$-boson, where it could be well reconstructed.
Denoting $\sigma_{S}$ and $\sigma_{B}$ as the signal and background
cross sections in the selected mass bins, we obtain the estimations
for the statistical significance ($SS$) of the signal by assuming an
integrated luminosity of $L_{int}=10^{4}$ pb$^{-1}$ per year,

\[
SS=\sqrt{2L_{int}\epsilon[(\sigma_{S}+\sigma_{B})\ln(1+\sigma_{S}/\sigma_{B})-\sigma_{S}]}.\]

The results for the signal significance of $t'$ single production
are given in Table \ref{tab2}. While calculating the $SS$ values
we consider the leptonic decay of $W^{\pm}$ bosons with the branchings
via $W^{\pm}\to l^{\pm}\nu{}_{l}$, where $l^{\pm}=e^{\pm},\mu^{\pm}$
and we assume the $b$-tagging efficiency as $\epsilon=60\%$. From
Table \ref{tab2}, we see that single $t'$ quark can be observed
at the LHeC with a mass in the range of 300-800 GeV provided optimal
mixings with the other families are present.

\begin{table}[htbp!]
\caption{The total cross section of signal ($\sigma_{s}$) and background ($\sigma_{b}$)
for the subprocess $e^{+}p\rightarrow W^{+}b\bar{\nu}_{e}$ ($e^{-}p\rightarrow W^{-}\bar{b}\nu_{e}$)
and its corresponding statistical significance ($SS$). While calculating
the $SS$ values we assume the $W^{\pm}$boson decays leptonically.\label{tab2}}

\begin{tabular}{llll}
\hline
$m_{t'}$(GeV)  & $\sigma_{s}(\bar{\sigma}_{s})$(pb)  & $\sigma_{b}(\bar{\sigma}_{b})$(pb)  & SS($\bar{SS}$) \tabularnewline
\hline
300  & 1.607x10$^{1}$ (1.607x10$^{1}$)  & 9.012x10$^{-5}$ (9.010x10$^{-5}$)  & 4113.27 (4113.08)\tabularnewline
400  & 7.503x10$^{-2}$ (7.504x10$^{-2}$)  & 1.230x10$^{-4}$ (1.228x10$^{-4}$)  & 1605.23 (1606.17) \tabularnewline
500  & 3.012x10$^{-2}$ (3.012x10$^{-2}$)  & 1.163x10$^{-4}$ (1.164x10$^{-4}$)  & 544.23 (544.09) \tabularnewline
600  & 1.120x10$^{-2}$ (1.119x10$^{-2}$)  & 8.894x10$^{-5}$ (8.892x10$^{-5}$)  & 171.39 (171.37) \tabularnewline
700  & 3.845x10$^{-3}$ (3.846x10$^{-3}$)  & 5.997x10$^{-5}$ (5.991x10$^{-5}$)  & 49.15 (49.17) \tabularnewline
800  & 1.210x10$^{-3}$ (1.211x10$^{-4}$)  & 1.047x10$^{-4}$ (3.852x10$^{-5}$)  & 8.35 (12.36)\tabularnewline
\hline
\end{tabular}
\end{table}

\section{Conclusion}

In the presence of the fourth family quarks, the LHC will discover
them in pairs and measure their masses with a good accuracy. The
single production may provide a unique measurement of the family
mixing with the four families. The $t'$ quarks can also be produced
singly at the LHeC with large numbers if they have a large mixing
with the other families of the SM. We have explored the single
production of sequential fourth family $t'$ quarks in $ep$ collision
at the LHeC energy with $\sqrt{s}=1.4$ TeV. It is shown that the
LHeC can discover single $t'$ quark up to 800 GeV with the optimized
values of mixing parameters.

\end{document}